\RequirePackage{snapshot}
\documentclass{aa}
\usepackage[varg]{txfonts} 
\usepackage{pgfplots}
\usepgfplotslibrary{external}
\usepackage{graphicx}
\usepackage{grffile}
\usepackage{units}

\usepackage[utf8]{inputenc}

\usepackage{amsmath}
\usepackage{amssymb}

\usepackage{caption}
\usepackage{subcaption}
\usepackage{units}

\usepackage{natbib}
\bibpunct{(}{)}{;}{a}{}{,} 

\usepackage{multirow}
\usepackage{tabularx}

\usepackage{import}

\DeclareGraphicsExtensions{.pdf,.png,.jpg}

\renewcommand{\phi}{\varphi}

\newcommand{\bm}{\boldsymbol}
\newcommand{\dd}{\,\mathrm{d}}
\DeclareMathOperator{\arcsinh}{arcsinh}
\DeclareMathOperator{\erf}{erf}

\DeclareMathOperator{\Ei}{Ei}

\newcommand{\sfrac}[2]{#1/#2}
\newcommand{\pdiff}[2]{\frac{\partial #1}{\partial #2}}
\newcommand{\diff}[2]{\frac{\mathrm{d}#1}{\mathrm{d}#2}}

\newcommand{\figpgf}[5][1.0]{\begin{figure}
\centering

\renewcommand\pgfimage[2][NOSTD]{\includegraphics[##1]{#2##2}}\resizebox{#1\linewidth}{!}{\input{#2#3.pgf}}
\caption{#5}
\label{#4}
\end{figure}
}

\let\originaleqref\eqref
\renewcommand{\eqref}{Eq.~\originaleqref}

\author{M.~Jung\inst{\ref{HS},\ref{CAU}} 
\and T.~F.~Illenseer\inst{\ref{CAU}} 
\and W.~J.~Duschl\inst{\ref{CAU},\ref{ARIZONA}}}

\institute{
Institute for Theoretical Physics and Astrophysics, Kiel Astrophysics, Christian-Albrechts-University Kiel, Leibnizstraße 15, D-24118 Kiel, Germany\label{CAU}
\and
Current affiliation: Hamburger Sternwarte, Universität Hamburg, Gojensbergweg 112, D-21029 Hamburg, Germany, \\\email{manuel.jung@hs.uni-hamburg.de}\label{HS}
\and
Steward Observatory, The University of Arizona, Tucson, AZ 85721, United States\label{ARIZONA}
}

\newcommand{\refree}[1]{{#1}}

\title{Multi-scale simulations of black hole accretion in barred galaxies: Numerical methods and tests.}


\abstract{
Due to the non-axisymmetric potential of the central bar, barred spiral galaxies form, in addition to their characteristic arms and bar, a variety of structures within the thin gas disk, like nuclear rings, inner spirals and dust-lanes. In this first of two papers, we present a method to accurately simulate the gas flow within the galactic plane in the 2D finite volume software package \textsf{FOSITE}, which solves the transport equations for mass, momentum and energy, and apply it to this class of objects. To this extent, we introduced a new transport scheme for angular momentum and a very efficient pseudo-spectral Poisson solver. Moreover, we provide a simple and generally applicable method of how to take care of gravity in the energy equation.
}

\begin{document}
\maketitle

\section{Introduction}
The $M$-$\sigma$-relation \citep{2000ApJ...539L...9F,2000ApJ...539L..13G,2011Natur.480..215M}
\begin{equation}
M \approx \unit[1.9\cdot 10^8 \left(\frac{\sigma}{\unit[200]{km/s}}\right)^{5.1}]{M_\odot},
\end{equation}
with $M$ the mass of the central black hole and $\sigma$ the bulge velocity dispersion, suggests a strong connection between black hole and evolution of the bulge.
This correlation is unexpected, because the related length scales differ by orders of magnitudes. 
Black hole accretion disks release high amounts of energy and are already at comparatively moderate accretion rates very luminous \citep{frank2002accretion}. These accretion rates can be deduced from the luminosity of active galactic nuclei. It becomes apparent that even in the local universe a broad spectrum of accretion rates spanning $\unit[10^{-5\dots 0}]{M_\odot/yr}$ is measured \citep{2012NewAR..56...93A}. \\
While the origin of the black hole-bulge connection remains generally unclear, galaxy merger \citep{milosavljevic_formation_2001}, AGN outflow \citep{king_agn-starburst_2005,king_powerful_2015} or stellar winds \citep{cid_fernandes_starbursts_2003} seem to be connected to it. It is unknown, to which extent these processes are applicable to other galaxies and how their effect on the collective growth operates.
The accretion of gas onto the black hole can only be explained by the transport of matter from much larger (kiloparsec) scales into the direct vicinity of the black hole and its gravitational sphere of influence ($\unit[<0.1]{pc}$). In doing so the gas not only has to loose nearly all of its angular momentum, but can also be subject to star formation by local gravitational collapse. This may suppress some of the gas flow directed into the central region. Recent high resolution observations \citep{2015ApJ...799...11X,2015ApJ...806L..34F,salak_gas_2016} slowly offer insight into the detailed generation of structures in the center of galaxies. Inner structures on the scale of $\unit[100]{pc}$ can be resolved e.g., clumpy nuclear rings, which show $1000$ times enhanced star formation rates compared to their surroundings \citep{2015ApJ...799...11X} or spiral arms spread from the center, which show gas movement towards the center \citep{salak_gas_2016}.\\
Often simulations of barred galaxies employ smoothed particle hydrodynamics \citep[SPH;][]{2002astro.ph..2004S,patsis_sph_2000,shlosman_nested_2002,ann_formation_2005}, amongst others because of the easy application and the almost exact conservation of angular momentum. A typical shortcoming of this method is the lack of resolution in the galactic center, as well as poor shock resolution, both of which, however, is important to study accretion flow onto a central black hole. Therefore in this work we use a grid-based hydrodynamic simulation code, which allows for static mesh refinement in the central region and naturally resolves shocks well. The underlying numerical scheme has been improved to ensure exact conservation of angular momentum in a rotating frame of reference, to remove any shortcomings compared to SPH.\\
To execute accurate measurements of the black hole accretion rate in the context of a bared galaxy, we developed a new numerical algorithm as part of the astrophysical software package \textsf{FOSITE} \citep{2009CoPhC.180.2283I}, which can solve systems of non-linear hyperbolic conservation equations. Hereby several computational challenges have arised, which are more generally important, not only with respect to the proposed galaxy simulations. The purpose of this paper is to introduce the new conservative transport scheme for angular momentum, the spectral self-gravity solver and the treatment of gravitational energy. The simulation of the gas flow in barred galaxies will be discussed in the second paper.\\
The paper is structured as follows. In section 2 we describe a method to ensure conservation of angular momentum, in section 3 a spectral self-gravity solver and in section 4 the correct treatment of gravitational energy. Section 5 is dedicated to tests of these new methods. In section 6 we conclude on our results and present an outlook on the second paper.

\section{Angular momentum conservation in a rotating frame of reference}
Angular momentum is one of the central conservation quantities in physics.
In accretion disk physics it is of particular importance, since matter which is accreted onto the central object has to get rid if nearly all of its angular momentum. 
Some matter gains a considerable amount of angular momentum and migrates to outer disk parts.
One of the central questions is the cause of this angular momentum and mass redistribution.
To gain detailed knowledge of these processes, numerical inaccuracies have to be ruled out.
Each numerical algorithm is an approximation of the analytical equations, which it is based on.
Therefore small errors in the calculation are inevitable.
The trick is to control the error and confine it to some well known limit.
Here we want to recover the angular momentum conservation, which is guaranteed by the underlying hydrodynamical equations.\\
If a rotating frame of reference is taken into account, special considerations are required. A naive implementation of the resulting fictitious forces leads to a exceptionally bad angular momentum conservation \refree{\citep{1998A&A...338L..37K}}. Hence it requires the application of particular techniques to improve the conservation properties in this case.\\
Vectorial conservation laws generate, if curvilinear orthogonal coordinates are chosen, in general geometrical source terms. The cause of this is the tensor divergence (c.f. \cite{2009CoPhC.180.2283I}), which includes derivations of the basis vectors. A naive implementation would destroy the conservation properties of the vectorial conservation law.\\
As a generalization of a rotating reference frame general solenoidal background velocity fields are considered.
In polar coordinates the transport of such velocity fields can be solved by means of a separate linear advection.
Especially for quasi stationary problems the timestep limitation as implied by the Courant–Friedrichs–Lewy condition \citep{courant_uber_1928} of the typically supersonic rotating accretion disk can be considerably lifted, since the linear advection is unconditionally stable \citep{2012A&A...545A.152M}.
The original idea for this method has been proposed by \cite{2000A&AS..141..165M} and is therefore often called FARGO advection. Several other astrophysical simulation codes \citep{2009NewA...14...71M,skinner2010athena,johnson2008orbital,2012A&A...545A.152M} have adopted the technique, which in principal accords to a moving mesh.
But the mesh is only moved by multiples of complete cells and therefore the geometry remains unchanged.
Since the emergence of the idea further developments of mixed Euler-Lagrangian methods have been proposed, e.g. moving annuli \citep{2010MNRAS.401..791S,Duffell2012} or moving meshs of arbitrary geometry \citep{Duffell2011}. However these methods pose their own problems like higher complexity of the algorithm, a tremendous computational demand and grid noise \citep{Duffell2012}.\\
The common equations describing the time evolution of a compressible inviscid fluid are given by the continuity equation for the density $\rho$
\begin{equation}\label{eq:kontinuitaetsgleichung}
\frac{\partial \rho}{\partial t} + \nabla \cdot \left(\rho \bm{v}\right) = 0,
\end{equation}
with the gas speed $\bm{v}$ and the momentum equation
\begin{equation}
\frac{\partial \left(\rho\bm{v}\right)}{\partial t} + \nabla \cdot \left(\rho\bm{v} \otimes \bm{v} + \mathbb{I}p\right) = 0.
\end{equation}
The energy equation describes the conservation of total energy, that is to say the sum of internal and kinetic energy $E = \frac{p}{\gamma-1} + \frac{1}{2}\rho\bm{v}^2$.
\begin{equation}\label{eq:energiegleichung}
\frac{\partial E}{\partial t} + \nabla \cdot \left(\left(E+p\right)\bm{v}\right) = 0
\end{equation}
If these equations are specialized to polar coordinates in a rotating reference frame, geometrical and fictitious source terms arise. Without special consideration of these in the numerical scheme, conservation of angular momentum and energy is lost.\\
We will now introduce a conservative implementation of the source terms.
The analytical derivation is similar to the method of \cite{2006A&A...450.1203P} and \cite{2012A&A...545A.152M}.
The system can also be used in other coordinates, which are useful for rotating fluids and reference systems. 
\cite{2012A&A...545A.152M} show derivations for spherical and cylindrical coordinates as well as the shearing box \citep{1995ApJ...440..742H}.\\
If $\bm{v}$ is the inertial velocity field and $\bm{w}$ a solenoidal vector field, e.g. the local velocity of a rotating reference frame $\bm{w} = \bm{\Omega} \times \bm{s}$.
Then $\bm{u} = \bm{v} - \bm{w}$ is called residual velocity, e.g. velocity in the rotating reference frame.
Analogous the residual energy density $E'$ is defined, which depends only on the residual velocity $\bm{u}$, but not on the inertial velocity:
\begin{equation}
 E' = \frac{p}{\gamma -1} + \frac{1}{2}\rho\bm{u}^2.
\end{equation}
These definitions allow us to transform the system eq. (\ref{eq:kontinuitaetsgleichung})~--~(\ref{eq:energiegleichung}) into
\begin{alignat}{3}
\frac{\partial \rho}{\partial t} &+ \nabla \cdot \left(\rho \bm{u}\right) &&+ \bm{w}\cdot\nabla\rho &&= 0\label{eq:pluto1}\\
\frac{\partial \left(\rho\bm{u}\right)}{\partial t} &+ \nabla \cdot \left(\rho\bm{u} \otimes \bm{u} + \mathbb{I}p\right) &&+ \bm{w}\cdot\nabla\left(\rho\bm{u}\right) &&= -\rho\bm{v}\cdot\nabla\bm{w}\label{eq:pluto2}\\
\frac{\partial E'}{\partial t} &+ \nabla \cdot \left(\left(E'+p\right)\bm{u}\right) &&+ \bm{w}\cdot\nabla E' &&= -\rho\bm{u}\cdot\left(\bm{v}\cdot\nabla\bm{w}\right)\label{eq:pluto3}.
\end{alignat}
We extend the original scheme by \citep{2012A&A...545A.152M} to arbitrary radial scalings in polar coordinates. It is therefore appropriate to use the following modified divergence operator in the azimuthal momentum equation to eliminate fictitious forces:
\begin{equation} 
\nabla_R\bm{F} = \frac{1}{h_\xi h_\phi^2}\pdiff{}{\xi}\left(h_\phi^2 F_1\right) + \frac{1}{h_\xi h_\phi}\pdiff{}{\phi}\left(h_\xi F_2\right).
\end{equation}
\refree{The resulting equations and metric coefficients in polar coordinates}
\begin{equation}
\bm{x} = r\left(\xi\right)\begin{pmatrix}\cos\left(\phi\right)\\ \sin\left(\phi\right)\end{pmatrix}
\end{equation}
using a monotonic differentiable radial scaling function $r\left(\xi\right)$ can be found in appendix~\ref{sec:polarcoo}.\\
The equations \eqref{eq:pluto1} - \eqref{eq:pluto3} (and \eqref{eq:plutopolar1} - \eqref{eq:plutopolar4}) are of the type
\begin{equation}
\pdiff{q}{t} + \nabla\cdot \bm{F}_q + \bm{w}\cdot\nabla q = S_q.
\end{equation}
Using operator splitting \citep{strang_construction_1968,mclachlan_splitting_2002} the linear transport term $\bm{w}\cdot\nabla q$ can be removed from the rest of the equation. This yields two differential equations
\begin{alignat}{2}
\pdiff{q}{t} &+ \nabla \cdot \bm{F}_q &&= S_q\\
\pdiff{q}{t} &+ \bm{w}\cdot\nabla q &&= 0,
\end{alignat}
in which the second equation accounts for the linear transport along the direction of $\bm{w}$. In this case $\bm{w}$ points in the direction $\bm{e}_\phi$ so that this equation can be solved numerically by shifting $q$ along the $\phi$ direction. This is done by an integer translation and a simple flux transport method for the remaining part of the translation velocity $\bm{w}$. Overall an implementation of the original \textsf{FARGO} scheme is obtained.
$\bm{w}$ is not restricted to the mean azimuthal velocity, but can accord to e.g. the velocity field of a rigid body rotation:
\begin{equation}
\bm{w} = r\Omega\bm{e}_\phi.
\end{equation}
If this is the case, the linear transport can be neglected, since it would only rotate the mesh entirely. There is no shear in the rotation motion of the mesh cell rings. Therefore this is equivalent to a rotating frame of reference with the angular velocity $\Omega$. \textsf{FOSITE} implements both cases so that hybrid forms 
\begin{equation}
\bm{w} = \left(\tilde{w} + r\Omega\right)\bm{e}_\phi
\end{equation}
are possible. The linear transport is only enabled for $\tilde{w}\neq 0$.\\
In case of a finite volume scheme the change of a quantity $u$ in the cell $(i,j)$ is defined by the flux of this quantity over the cell boundaries and external sources. Since \textsf{FOSITE} implements a semi-discrete scheme, the spatial dimensions are discretized, while the time dimension stays independent from this discretization. If we call the numerical fluxes in the different directions at the cell boundaries $\mathcal{F}$ and $\mathcal{G}$, as well as $S$ the external sources, we can write down the change of the quantity $u$ in the cell $(i,j)$ \citep{2009CoPhC.180.2283I}:
\begin{equation}\label{eq:flussschema}
\diff{u_{i,j}}{t} = -\frac{\mathcal{F}_{i+\frac{1}{2}} -\mathcal{F}_{i-\frac{1}{2}}}{\Delta V_{i,j}}-\frac{\mathcal{G}_{j+\frac{1}{2}} -\mathcal{G}_{j-\frac{1}{2}}}{\Delta V_{i,j}} + \left\langle S \right\rangle_{D_{i,j}}.
\end{equation}
In case of pure conservation laws $S=0$ holds.\\
Our new scheme will be conservative, if the considered quantity is exclusively modified by cell boundary fluxes. The goal is to reformulate the new source terms, which are generated by the separation of the $\bm{w}$, in terms of fluxes over the cell boundaries. The ordinary physical fluxes are identified by \citep{2009CoPhC.180.2283I}
\begin{align}
\bm{F}_\rho &= \rho\bm{u}\\
\bm{F}_{m_\xi} &= \rho u_\xi \bm{u} + p \bm{e}_\xi\\
\bm{F}_{m_\phi} &= \rho u_\phi \bm{u} + p \bm{e}_\phi\\
\bm{F}_E &= \left(E' + p\right)\bm{u}.
\end{align}
These are subject to the classical numerical flux calculations, e.g. Riemann solvers \citep{Harten1983,Toro1994,Roe1981} or the Kurganov-Tadmor scheme \citep{Kurganov2000}. The equations \eqref{eq:plutopolar1} - \eqref{eq:plutopolar4} are now implemented similar to the form of equation \eqref{eq:flussschema}:
\begin{align}
\diff{\rho}{t} &= -\left\langle\nabla \cdot \bm{F}_\rho\right\rangle\\
\diff{\left(\rho u_\xi\right)}{t} &= -\left\langle\nabla\cdot \bm{F}_{m_\xi}\right\rangle + \rho v_\phi^2 c_{\phi\xi\phi}\\
\diff{\left(\rho u_\phi\right)}{t} &= -\left\langle\nabla_R\cdot\left(\bm{F}_{m_\phi} + w\bm{F}_\rho\right)\right\rangle + w\left\langle\nabla \cdot \bm{F}_\rho\right\rangle\\
\diff{E'}{t} &= -\left\langle\nabla\cdot\left(\bm{F}_E + w\bm{F}_{m_\phi} + \frac{w^2}{2} \bm{F}_\rho\right)\right\rangle + w \left\langle\nabla_R\cdot\left(\bm{F}_{m_\phi} + w\bm{F}_\rho\right)\right\rangle\nonumber\\
&\phantom{{}=}+ \frac{w^2}{2}\left\langle\nabla\cdot\bm{F}_\rho\right\rangle,
\end{align}
in doing so terms like $\langle \nabla\cdot \bm{F} \rangle$ have to be replaced by quotient of a flux difference and the volume of a cell from equation \eqref{eq:flussschema}. Unfortunately the derivation in \cite{2012A&A...545A.152M} included minor errors, which is why we repeat it here. Note that all terms can be calculated using the original physical fluxes and the right hand side of the azimuthal momentum equation includes the complete right hand side of the continuity equation, as well as the right hand side of the energy equation includes the complete right hand side of the azimuthal momentum equation.

\section{Self-gravity solver in polar coordinates}
Massive accretion disks produce sufficient gravitational accelerations so that the gas mass dominates the overall gravitational potential, at least in some regions. Moreover, under certain conditions the accretion disk becomes gravitational unstable \citep{1964ApJ...139.1217T}.\\
The Poisson equation
\begin{equation}\label{gl:poisson}
\nabla^2\Phi = 4\pi G \rho
\end{equation}
is applied for a mass distribution with mass density $\rho$, $G$ the gravitational constant and $\Phi$ the gravitational potential. In this work we assume that $\Phi$ complies with the boundary condition of a vanishing potential at infinity
\begin{equation}
\lim_{\left|\bm{x}\right|\rightarrow \infty}\Phi\left(\bm{x}\right) = 0.
\end{equation}
Generally the Poisson equation can be solved either by a finite difference method \citep{leveque2007finite} or through direct integration. The first class of methods directly solves the Poisson equation \eqref{gl:poisson} and requires suitable boundary conditions, which can be quite difficult.\\
In case of the direct integration method, we have to solve the volume integral
\begin{equation}\label{gl:poissonint}
\Phi\left(\bm{x}\right) = -G\int_V\frac{\rho\left(\bm{x}'\right)}{\left|\bm{x}-\bm{x}'\right|}d\bm{x}'.
\end{equation}
Here the boundary conditions are not a problem, since the integral is only calculated inside the computational domain. However a difficulty arises from the evaluation at the singular point $\bm{x}'\rightarrow \bm{x}$. Additionally, integral methods a typically computationally expensive. In the following we introduce a direct integration implemented with Fourier transformations, which can partially be cached and reused. Singular points can be avoided through introduction of a staggered mesh for the Fourier space evaluations. The method is based on  \cite{2006ApJ...645..506C} and \cite{2009ApJS..181..244L}, but differs in the choice of the secondary grid and is generalized to polar coordinates with an arbitrary radial scaling.\\
We assume for simulations of accretion disks in 2D polar coordinates that they are geometrically thin. Therefore, the scale height $H$ of a accretion disk fulfills
\begin{equation}
\frac{H\left(r\right)}{r} \ll 1.
\end{equation}
Additionally we assume that the scale height depends only on the radial coordinate but not on the azimuthal coordinate $\phi$ or the time $t$:
\begin{equation}
H = H\left(r\right).
\end{equation}
Then using cylindrical coordinates the vertical structure can be described by a function $Z\left(r,z\right)$. The volume density $\rho$ is therefore decomposed as a product of surface density $\Sigma$ and vertical structure $Z$:
\begin{equation}
\rho\left(t,r,\phi,z\right) = \Sigma\left(t,r,\phi\right)Z\left(r,z\right).
\end{equation}
Using this in \eqref{gl:poissonint} and solving for $z=0$ provides:
\begin{align}\label{eq:poissonintmithoehe}
\Phi\left(t,r,\phi\right)&=\int_{r_\mathrm{min}}^{r_\mathrm{max}}\int_{0}^{2\pi}\Sigma\left(t,r',\phi'\right)\nonumber\\
&\phantom{{]}=}\cdot\int_{-\infty}^{\infty}-\frac{G Z\left(r',z'\right)}{\sqrt{r^2+r'^2-2 r r'\cos\left(\phi-\phi'\right)+z'^2}}\nonumber\\
&\phantom{{]}=\cdot\int_{-\infty}^{\infty}}\dd z' \sqrt{g}\dd\phi'\dd r'
\end{align}
Here $\sqrt{g}$ is the Jacobian determinant. The last integral is the Green's function
\begin{equation}
\mathcal{G}\left(r,r',\phi-\phi'\right) = \int_{-\infty}^{\infty}-\frac{Z\left(r',z'\right)}{\sqrt{r^2+r'^2-2 r r'\cos\left(\phi-\phi'\right)+z'^2}}\dd z'.
\end{equation}
Since the vertical structure $Z$ is assumed to be time independent, it can be calculated once at the beginning of the simulation. All following vertical structures have known analytic solutions to the integral. The simplest case is a razor thin disk
\begin{align}
Z\left(r,z\right) &= \delta\left(z\right), &
\mathcal{G}\left(r,r',\phi-\phi'\right) &= \frac{1}{\sqrt{r^2+r'^2-2 r r' \cos\left(\phi-\phi'\right)}}.
\end{align}
A geometrical thin, vertically isothermal, non self-gravitating accretion disk features as vertical structure a Gaussian distribution \citep{lynden-bell_galactic_1969}
\begin{equation}
Z\left(r,z\right) = \frac{1}{\sqrt{2\pi}H\left(r\right)}\exp\left(-\frac{z^2}{2 H^2\left(r\right)}\right).
\end{equation}
This applies also to vertically isothermal self-gravitating accretion disks \citep{2015MNRAS.450..691I}. Using a scale height $H\left(r\right)$ we can calculate the vertical structure and the Green's function:
\begin{equation}
\mathcal{G}\left(r,r',\phi-\phi'\right) = -\frac{e^{R^2/4}K_0\left(R^2/4\right)}{\sqrt{2\pi}H\left(r'\right)},
\end{equation}
using $R^2=\left(r^2+r'^2-2 r r' \cos\left(\phi-\phi'\right)\right)/H^2\left(r'\right)$ and $K_0$ the modified Bessel function of the second kind (see addendum~\ref{sec:besselk0}).
Employing the definition
\begin{equation}
I\left(r,r',\phi-\phi'\right) := 2\pi G r' \mathcal{G}\left(r,r',\phi-\phi'\right)
\end{equation}
we can rewrite the integral \eqref{eq:poissonintmithoehe} as:
\begin{equation}
\Phi\left(t,r,\phi\right) = \int_{r_\mathrm{min}}^{r_\mathrm{max}}\frac{1}{2\pi}\int_0^{2\pi}\Sigma\left(t,r',\phi'\right) I\left(r,r',\phi-\phi'\right)\dd r'\dd \phi'.
\end{equation}
The second integral is the convolution of $\Sigma$ and $I$. If we define $\hat{\Sigma}$ and $\hat{I}$ as the Fourier transforms of $\Sigma$ and $I$ (see addendum~\ref{sec:fouriertrafo}) and use the convolution theorem we obtain the Fourier transform of the gravitational potential
\begin{equation}\label{eq:mittelpunktint}
\hat{\Phi}_m\left(t,r\right) = \int_{r_\mathrm{min}}^{r_\mathrm{max}} \hat{\Sigma}_m\left(t,r'\right) \hat{I}_m\left(r,r'\right)\dd r' \qquad m\in\left[-\infty,\infty\right].
\end{equation}
This integral can easily calculated with the rectangle method. To avoid the divergence of the solution for $r\rightarrow r'$, we use different grids for $r$ and $r'$. $r'$ is evaluated at the cell centers, since $\Sigma$ is also defined at these points and therefore unnecessary reconstruction is prevented. $r$ is defined at the cell boundaries. This choice alleviates the computation of the gravitational acceleration. The following steps have to be carried out for every source term evaluation:
\begin{enumerate}
	\item Calculation of the Fourier transform of $\Sigma$
	\item Evaluation of the integral \eqref{eq:mittelpunktint} using the rectangle method
	\item Inverse Fourier transform of $\hat{\Phi}$
\end{enumerate}
As preparation we have to calculate the Fourier transform of $I$ before the simulation starts. The actual algorithm at every timestep consists only of two Fourier transformations and the summation of products (see equation~\eqref{eq:mittelpunktint}). This method is very fast, since for the Fourier transformation very fast and efficient algorithms are available. Typically the computation of the gravitational acceleration due to self-gravity uses about one third of the wall time of a simulation. Since 1D Fourier transformations are not efficiently parallizable, we choose to only do domain decomposition along the radial direction in rings. \refree{Furthermore, this simplifies the implementation of the orbital advection scheme, which otherwise might need additional MPI communication.}

In case of very high resolution simulations low frequencies in the Fourier space may be neglected for the radial summation. This does not only reduce the amount of summation operations in the radial direction, but also the amount of communication. Tests show that the truncation error for neglecting half of the modes is typically in the lower single-digit range \cite{2009ApJS..181..244L}. This scheme is also implemented in \textsf{FOSITE}, but it is not used in the following simulations, since at the used resolutions the communication costs are still manageable.\\
The gravitational acceleration can be obtained computing the gradient of the potential
\begin{equation}
\bm{g} = -\nabla\Phi,
\end{equation}
which is done using finite difference approximations for the acceleration at cell centers $(i,j)$ according to
\begin{align}
g_r\left(r_i,\phi_j\right) &= \frac{\Phi\left(r_{i+\frac{1}{2}},\phi_j\right)-\Phi\left(r_{i-\frac{1}{2}},\phi_j\right)}{\Delta r},\\
g_\phi\left(r_i,\phi_j\right) &= \frac{\Phi\left(r_{i+\frac{1}{2}},\phi_{j+1}\right)+\Phi\left(r_{i+\frac{1}{2}},\phi_{j-1}\right)-\Phi\left(r_{i-\frac{1}{2}},\phi_{j-1}\right)}{4r_i \Delta\phi}.
\end{align}

\section{Treatment of gravity in the energy equation}
Accretion disks are approximately in a balanced state of gravitational force, pressure gradient and centrifugal force \citep{1973A&A....24..337S,1981ARA&A..19..137P}.
The pressure gradient in the radial direction is typically small compared to the gravitational forces and can therefore be neglected.
If gravitational and centrifugal forces exactly cancel each other, the system is in a state of equilibrium.
Therefore, if the radial velocity is zero, the radial momentum flux vanishes.
The difficulties arising from a large imbalance of internal and total energy were already noticed by \cite{einfeldt1991godunov}. Nonetheless their method demands a transformation of the complete numerical scheme and violates its numerous different requirements and capabilities.\\
We propose a new method, which has the advantage to be applicable for self-gravitating potentials. The elegant alternative method described by \cite{jiang2013new} demands a complicated modification of the conservative variables in the energy equation and thus is not readily suitable for the reformulated rothalpy equation. In contrast the new method in this work enables conservative energy transport in a rotating reference frame with self-gravitation. If the common formulation of the energy equation
\begin{equation}
\pdiff{E}{t} + \nabla \cdot \left(\left(E+p\right)\bm{v}\right) = -\rho\bm{v}\cdot \nabla\Phi
\end{equation}
is used for accretion disks, a huge amount of kinetic energy has to be transported in the azimuthal direction. This result of the divergence term is mostly canceled by the potential energy source term with no net change in total energy. In the most simple case the axisymmetric potential as well as the absolute value of the velocity field feature a radial gradient. The radial gradient of the gravitational potential generates a source in the energy equation, which has to be balanced by the radial velocity gradient on the left hand side. It is extremely difficult to achieve a well balanced solution, because the flux on the left hand side is subject to reconstruction and numerical flux calculation.\\
Similar to the well balanced formulation of the source terms to achieve exact angular momentum conservation, the potential energy is reformulated in an analytical equivalent form, in such a way as to transport the gravitational energy together with the matter.
Since the gravitational and kinetic energy have a similar absolute value, but opposite signs, they cancel each other approximately. Therefore the numerical fluxes only transport the difference to the balance of gravitational and centrifugal forces. This increases the accuracy, since minor deviations do not get lost because of truncation errors \citep{atkinson2008introduction}.
The gravitational source term in the energy equation is expanded by
\begin{equation}
-\rho\bm{v} \cdot \nabla\Phi = -\nabla\cdot\left(\rho\Phi\bm{v}\right) + \Phi\nabla\cdot\left(\rho\bm{v}\right).
\end{equation}
Than the energy equation can be reformulated:
\begin{equation}
\pdiff{E}{t} = -\nabla\left(\left(E+p+\rho\Phi\right)\bm{v}\right) + \Phi \nabla\cdot\left(\rho\bm{v}\right).
\end{equation}
Now the gravitational energy is a transport term and is not handled as a common source term.
The implementation follows the same concept as for the balanced formulation of the angular momentum source terms.
First, the numerical fluxes \refree{of eq.~\ref{eq:energiegleichung}} are calculated, than the gravitational energy is added and the spatial derivative is computed.
The second term on the right hand side is handled by multiplication of the continuity equation with the potential. Thus the total energy conservation is considerably improved.
All of the following simulations with energy equation and gravitational potentials are only possible because of this method. Otherwise errors are introduced typically near massive central objects in the manner of numerical instabilities, respectively negative pressure. In this case the errors in calculating internal, kinetic and potential energy are simply too large. Without the modified transport including angular momentum conservation these changes to the energy equation are from experience nonessential. Then the intrinsic dissipation of the numerical scheme is adequate to assure the stability of the solution. In the next section a simple example will test the reformulated energy equation.

\section{Test problems and results}
\subsection{Isothermal and isentropic vortex test}\label{sec:isenvortex}
The angular momentum conservation scheme is now tested with the setup of an isentropic and locally isothermal vortex. The isentropic vortex setup is based on \cite{1999JCoPh.150..199Y}. The main idea is the same for both equations of state: The centrifugal forces of a rotating fluid are balanced by a radial pressure gradient. The result is a stationary flow, which has to be preserved by the numerical scheme. The original setup by \cite{1999JCoPh.150..199Y} requires to solve the energy equation. We also approximate the stationary flow in a locally isothermal setup. \\
Let $\rho_\infty$ be the background density, $T_\infty$ the background temperature and $u_\infty$, $v_\infty$ the background velocities in Cartesian coordinates of the undisturbed fluid. Using dimensionless quantities
\begin{align}
\rho_\infty &= T_\infty = p_\infty = 1,\\
u_\infty &= v_\infty = 0
\end{align}
holds \citep{oertel_stromungsmechanik._2005}. Now we add the isentropic vortex as disturbance. An arbitrary velocity field can be chosen, which tends to zero as $r\rightarrow 0$ or $r\rightarrow \infty$. In addition the entropy shall be conserved ($\delta S=0$):
\begin{align}
\begin{pmatrix}\delta u\\ \delta v\end{pmatrix} &= \frac{\beta}{2\pi}\exp\left(\frac{1-r^2}{2}\right)\begin{pmatrix}-y\\x\end{pmatrix},\\
\delta T &= -\frac{\left(\gamma-1\right)\beta^2}{8\gamma\pi^2}\exp\left(1-r^2\right).
\end{align}
Here $\beta=5$ is the vortex strength, $\gamma = 1.4$ the adiabatic index and $r^2=x^2+y^2$.
Using
\begin{align}
\rho = \rho_\infty+\delta\rho, \quad &T = T_\infty + \delta T, \quad p = \rho T,\\
u = u_\infty+\delta u, \quad &v = v_\infty+\delta v
\end{align}
and the isentropic condition for an ideal gas $p/\rho^\gamma=p_\infty/\rho_\infty^\gamma=1$, we yield the analytic expression\footnote{The density solution has been corrected by a factor of $\sfrac{1}{\pi}$, which is missing in the derivation of  \cite{1999JCoPh.150..199Y}.} for stationary flow with respect to Cartesian coordinates:
\begin{align}
\rho &= \left(1-\frac{\left(\gamma-1\right)\beta^2}{8\gamma\pi^2}\exp\left(1-\gamma^2\right)\right)^{1/\left(\gamma-1\right)},\\
\begin{pmatrix}u\\v\end{pmatrix} &= \frac{\beta}{2\pi}\exp\left(\frac{1-r^2}{2}\right)\begin{pmatrix}-y\\x\end{pmatrix},\\
p &= \rho^\gamma.
\end{align}
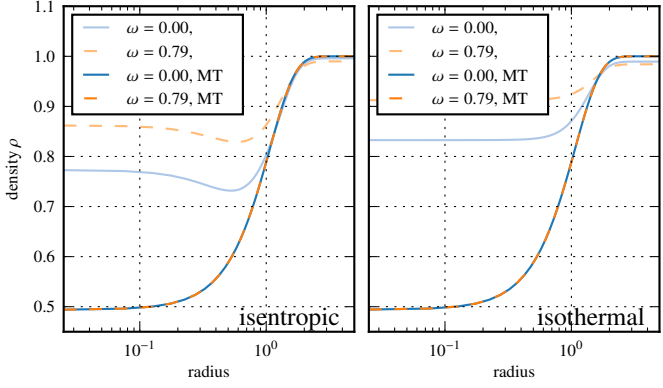
\figpgf{sims/figure/vortex2dpolar/}{density_2000}{vortex2dpolar_density_2000}{Radial density profile of the isentropic (left) and locally isothermal (right) vortex at the end of the simulation. Density and radius are dimensionless. The simulation without the modified transport scheme (pale blue and orange) show a strongly dissipative behavior and a progressive disintegration of the vortex. A rotating frame of reference (pale orange) accelerates this effect. If the modified transport scheme (MT, strong blue and orange) is used, the rotation speed of the reference frame is irrelevant. The vortex does not show any variation during the simulation time.}
If we use a locally isothermal equation of state, the density and velocity fields are defined as before. The local sound speed is a function of the density
\begin{equation}
c_s = \rho^{\sfrac{\left(\gamma-1\right)}{2}}.
\end{equation}
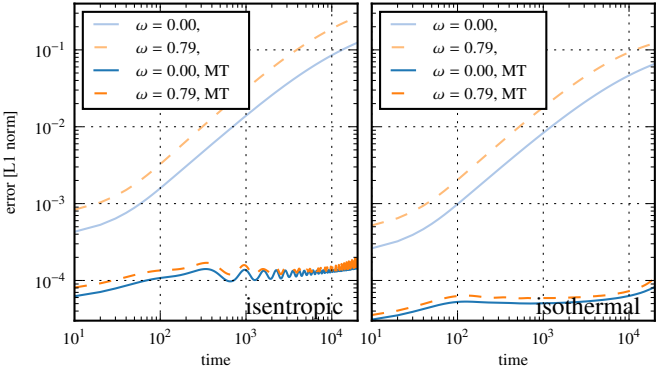
\figpgf{sims/figure/vortex2dpolar/}{L1_12}{vortex2dpolar_L1_12}{Time progression of deviation to the initial stationary state for the isentropic and locally isothermal vortices using both transport schemes. The time is dimensionless. Generally simulations without an rotating reference frame show a slightly smaller error. Without the modified transport the error is about $10\%$ at the end of the simulation, in contrast the modified transport (MT) shows an error of less than $2\cdot 10^{-4}$ most of the time.}
We execute simulations of the isentropic and locally isothermal vortex with and without angular momentum conservation. In addition we do these simulation in a rotating frame of reference with the angular velocity $\omega=0.79$, which corresponds to the maximum rotational velocity of the setups. At the radius of highest angular velocity the vortex has undergone about $4175$ orbits at the end of the simulation $t_\mathrm{sim}=2\cdot 10^4$. The polar grid spans a radial range from $R=0$ to $R=5$. Therefore we use \textsf{AXIS} conditions on the inner and \textsf{NO\_GRADIENTS} on the outer boundary. The grid resolution is $100\times 10$ and the primitive variables are reconstructed on the cell faces using the \refree{van Leer} slope limiter. The flux calculation is done with the Harten-Lax-van Leer-Contact scheme (abbr. HLLC, \citealt{Toro1994}).\\
Figure~\ref{vortex2dpolar_density_2000} shows the radial density of the isentropic, respectively locally isothermal of the simulation. The classic transport scheme is not able to preserve the vortex. After a short amount of time it has already lost most of its profile. On the contrary the modified transport scheme is able to preserve the vortex profile even for very long periods time.\\
Figure~\ref{vortex2dpolar_L1_12} shows the deviations from the initial configurations in the L1 norm as a function of time. The error of the simulation without modified transport is already higher than in the simulations with angular momentum conservation at any time. In the further progress the error for the classic scheme rises considerably, while it stays nearly constant for simulations with modified transport. Only at the end of the simulation some increase is noticeable.

\subsection{Density-potential pairs}
The first test of the new spectral self-gravity solver is the comparison of a density distribution with known analytic solution for the gravitational potential. The self-gravity module only calculates the gravitational potential of mass within the computational domain, but does no allow for boundary conditions, which account for changes of the gravitational potential due to material located outside the computational domain. Therefore density distributions should become negligible near the boundaries of the computational domain. The potential of an outside residing mass distribution can be modeled with other gravitational modules.\\
Useful potentials with this property are typically mass distributions, which are centered around a point in space and decline with increasing distance to this point. Such distributions are known for razor thin disks as well as disks with a Gaussian vertical structure.\\
In case of razor thin disks we define a around $(r_k,\phi_k)$ centered mass distribution
\begin{equation}
\Sigma_{\left(r_k,\phi_k\right)}\left(r,\phi\right)=\frac{1}{2\pi\sigma^2}\exp\left(-\frac{R_k}{\sigma}\right)
\end{equation}
with $\sigma$ a measure for the radial extend of the disk. $R_k$ identifies the distance of a point in space to the center of the distribution:
\begin{equation}
R_k = \sqrt{r^2+r_k^2-2 r r_k \cos\left(\phi-\phi_k\right)}.
\end{equation}
Such a mass distribution generates the potential \citep{2006ApJ...645..506C}:
\begin{equation}
\Phi_{\left(r_k,\phi_k\right)}\left(r,\phi\right) = -\frac{G}{\sigma}\left(I_0\left(y_k\right) K_1\left(y_k\right)-I_1\left(y_k\right) K_0\left(y_k\right)\right),
\end{equation}
with $y_k = R_k/\left(2\sigma\right)$. $I_n$ and $K_n$ denote the modified Bessel functions of the first respectively second kind of integer order (see adendum~\ref{sec:besselk0}).\\
In the following we setup a composition of three of these mass distributions at a resolution of $1024\times 3072$ in polar coordinates, similar to a test setup of \cite{2006ApJ...645..506C}. The solution for the gravitational potential is than the superposition of the solutions of each individual mass distribution. The surface density distribution is defined by
\begin{align}
\Sigma\left(r,\phi\right) &= 2\Sigma_{\left(1,10^{-3}\right)}\left(r,\phi\right) + 0.5\Sigma_{\left(1,\pi+10^{-3}\right)}\left(r,\phi\right)\nonumber\\
&\phantom{{}={}}+ \Sigma_{\left(0.9, \frac{3}{4}\pi\right)}\left(r,\phi\right).
\end{align}
The solution for the potential is:
\begin{align}
\Phi\left(r,\phi\right) &= 2\Phi_{\left(1,10^{-3}\right)}\left(r,\phi\right) + 0.5\Phi_{\left(1,\pi+10^{-3}\right)}\left(r,\phi\right) \nonumber\\
&\phantom{{}={}}+ \Phi_{\left(0.9, \frac{3}{4}\pi\right)}\left(r,\phi\right).
\end{align}
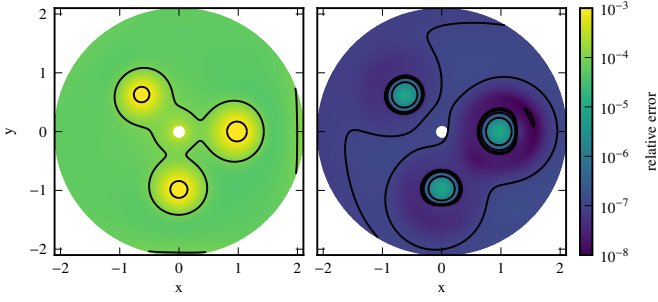
\figpgf{sims/figure/orbitingcylinders/}{relerror12}{relerror12}{Relative error of the numerical solution of the gravitational potential of flat cylinders (left) and cylinders with a vertical Gaussian distribution with varying masses.}
Figure~\ref{relerror12} (left) shows the relative error of the numerical solution in comparison to the analytical potential. The maximum error is less than $10^{-3}$ at the points of maximum density and is typically about $10^{-5}$ if there are no large density gradients.

Next we test the implementation of the Green's function for a vertical Gaussian distribution, using $\sigma=0.1$ as a measure for the scale height. A density distribution, which is concentrated around the $\left(r_k, \phi_k\right)$
\begin{equation}
\Sigma_{\left(r_k,\phi_k\right)}=\frac{1}{2\pi \sigma^2} \exp\left(-\frac{R_k^2}{2\sigma^2}\right),
\end{equation}
generates the gravitational potential
\begin{equation}
\Phi\left(r_k,\phi_k\right) = -\frac{1}{R_k}\erf\left(\frac{R_k}{\sqrt{2}\sigma}\right).
\end{equation}
Figure~\ref{relerror12} (right) shows the relative error, if we use the same points in space and masses as before. The vertical Gaussian mass distribution has a maximum relative error, which is smaller than $10^{-5}$ at the points of heighest density and elsewhere of the order $10^{-7}$.\\
The results of density-potential-tests of \cite{2006ApJ...645..506C} show a similar structure of the error distribution. The maximum relative error in the simulation with vertical Gaussian mass distribution is in both works equal ( $10^{-3}$). A razor thin disk has a maximum relative error in this work of $10^{-5}$ about two orders of magnitude lower than \cite{2006ApJ...645..506C} results. A more detailed comparison is difficult, since \cite{2006ApJ...645..506C}, e.g. do not document the resolution of the computational domain.

\subsection{Self-gravitating rotating cylinders}\label{sec:orbitingcylinders}
To test not only the gravitational potential calculations, but also the gravitational acceleration and the behavior in a time dependent simulation, we use a setup introduced by \cite{2006ApJ...645..506C}. In this simulation two cylinders with a Gaussian density distribution are generated, which exactly counterpart each other in the polar computational domain. A background velocity field moves the computational domain in a rigid rotation. The therefore generated centrifugal forces have to be compensated by self-gravitation of the mass distribution. This is a very demanding setup for the accuracy of the acceleration generated by the self-gravitation and the general angular momentum conservation.\\
The initial conditions are defined by the density distribution
\begin{align}
\rho\left(r,\phi\right) &= \frac{10^{-2}}{\pi\left(r_\mathrm{max}^2-r_\mathrm{min}^2\right)}+ 0.99\left(\frac{1}{2\pi\sigma^2}\exp\left(-\frac{R_1}{2\sigma^2}\right)\right. \nonumber\\
&\phantom{{}={}} \left.+ \frac{1}{2\pi\sigma^2}\exp\left(-\frac{R_2}{2\sigma^2}\right)\right),
\end{align}
and the pressure
\begin{align}
p\left(r,\phi\right) &= \frac{10^{-2}}{\pi\left(r_\mathrm{max}^2-r_\mathrm{min}^2\right)} + \frac{G}{2\pi\sigma^2}\left(\Ei\left(-\frac{R_1}{\sigma^2}\right) - \Ei\left(-\frac{R_1}{2\sigma^2}\right)\right.\nonumber\\
&\phantom{{}={}}\left.+ \Ei\left(-\frac{R_2}{\sigma^2}\right) - \Ei\left(-\frac{R_2}{2\sigma^2}\right)\right)
\end{align}
as well as the velocity field
\begin{equation}
\bm{v} = r\bm{e}_\phi.
\end{equation}
Here $\Ei$ denotes the exponential integral (see, e.g. \citealt{1965hmfw.book.....A}).\\
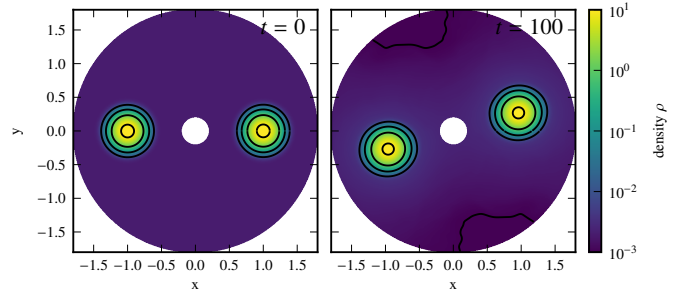
\figpgf{sims/figure/orbitingcylinders/}{density}{orbcyl_density}{Color map of the density for the rotating self-gravitating cylinders at the start and end of the simulation. The shape of the cylinder is very well conserved. The maximum density value decreases slightly.}
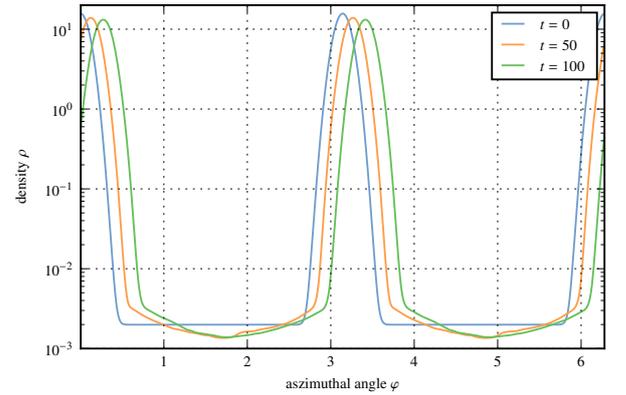
\figpgf[0.9]{sims/figure/orbitingcylinders/}{peaks}{orbcyl_peaks}{Azimuthal cut through the points of maximum density for different points in time in the simulation of rotating self-gravitating cylinders. The density maximum is very well conserved. In the area of artificial density cutoff, a adaptation process takes place.}
The simulation is executed on a polar grid with a resolution of $256\times 1024$ cells and a radial extent of $r\in\left[0.2,1.8\right]$. At these boundaries we use reflecting boundary conditions. The full set of Euler equations is closed by an isentropic equation of state with an adiabatic exponent of $\gamma=\sfrac{5}{3}$. For the flux calculation we use the Kurganov-Tadmor fluxes \citep{Kurganov2000} with linear reconstruction, slope limited by the \refree{van Leer} limiter \citep{vanLeer1974}. The simulations runs for $100$ orbits.\\
Figure~\ref{orbcyl_density} shows the initial condition and the final state of the density distribution. The cylinders circular shape is very well conserved. Figure~\ref{orbcyl_peaks} shows the density in azimuthal direction at the radius of maximum density for different points in time. The density decreases slightly, but again the shape is very well conserved. The difference of the solution at the transition to the background distribution is expected, since this is an artifical cutoff.\\
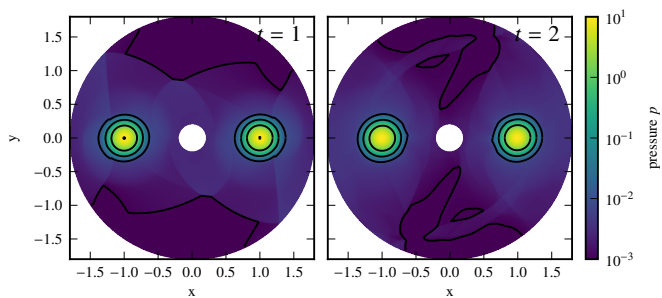
\figpgf{sims/figure/orbitingcylinders/}{pressure_12}{orbcyl_pressure_12}{Pressure color map of the simulation of rotating self-gravitating cylinders. A shock wave is visible, which is generated directly after the initial state, since it is not completely in equilibrium.}
Figure~\ref{orbcyl_pressure_12} shows the pressure in the first two outputs $t\in\{1,2\}$ at the beginning of the simulation. Apparently the initial condition is not in perfect equilibrium, since both cylinders eject some of their energy as a shock wave. But this is for the following examination of the conservation properties not of particular interest. Still it explains, why the decrease in peak density is stronger in the first half of the simulation.\\
The simulation loses in a rotating reference frame after $100$ orbits of the cylinders $0.06\%$ of the total angular momentum. Also, \cite{2006ApJ...645..506C} conserve the shape of the cylinders relatively well, but loose in the course of $16$ orbits already $2\%$ of the total angular momentum. Again their grid resolution is unknown, but they need $5\cdot 10^4$ timesteps for $16$ orbits, in contrast to our simulation, which needs less than $5\cdot 10^5$ for $100$ orbits. Therefore, we deduct a about $200\times$ better angular momentum conservation per timestep for our scheme with angular momentum conservation.

\subsection{Vortex transport in a disk with keplerian rotation}
The goal of the following test is to check the ability of the numerical scheme to preserve an anticyclonic vortex embedded in a Keplerian disk in a long-term simulation. If the preservation of a vortex is sufficiently strong, one can act on the assumption that such a vortex can be formed due to outer influences. An extensive description of the test can be found in its original source \cite{2007A&A...475...51B}. Further data for comparison can be found in \cite{2012A&A...545A.152M}.\\
Initially the Keplerian background velocity is defined as  $\rho_\infty=1$ and the pressure as $p_\infty=1/\left(\gamma M^2\right)$, using the Mach number $M=10$ at radius $r=1$. Then we add a circular vortex to the background flow:
\begin{equation}
\begin{pmatrix}\delta v_r\\\delta v_\phi \end{pmatrix} = \kappa \exp\left(-\frac{x^2+y^2}{h^2}\right)\begin{pmatrix}\cos\phi & \sin\phi\\-\sin\phi & \cos\phi\end{pmatrix}\begin{pmatrix}-y\\x\end{pmatrix}.
\end{equation}
Here $x$ and $y$ are Cartesian coordinates with respect to the center of the vortex $(x_0,y_0)$, $\kappa=-1$ the vortex strength and $h=1/(2M)$ a dimensionless scale parameter.\\
We run simulations at resolutions of $256\times 1024$, $512\times 2048$ and $1024\times 4096$ with Kurganov-Tadmor (abbr. KT, \citealt{Kurganov2000}) and HLLC fluxes. We use the same equation of state, reconstruction method and limiter as in the last section. The radial extent of the computational domain is $r\in\left[0.4,2.0\right]$. For the reflecting boundary conditions we choose a Keplerian continuation of the azimuthal velocity. The simulations run until the vortex reached $30$ orbits.\\%
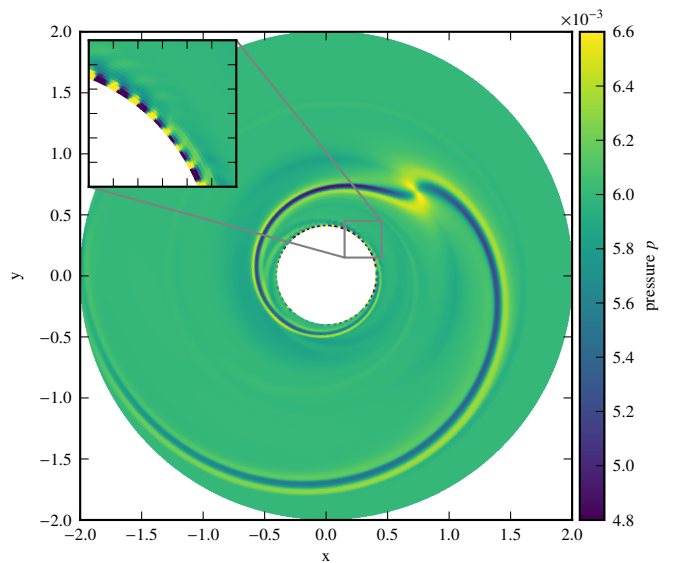
\figpgf{sims/figure/keplerianvortex/30_0256/}{pressure_zoom}{kepvort_noenergy}{Pressure of a simulation with HLLC fluxes and $256$ radial cells without the modified gravitational energy transport scheme. A strong numerical instability emerges already after two orbits, shortly after the inner spiral arm or more general a non-axisymmetric perturbation hits the inner boundary. The consequences are negative pressure and therefore abort of the simulation.}
In the course of all simulations the initially circular vortex is sheared slightly along the azimuthal direction. At the same spiral density waves are generated, which slowly inch towards the domain boundaries, where they are reflected and as a consequence cross themselves. Depending on the chosen flux type and resolution, the vortex will be dissolved and the spiral arms may not be noticeable anymore.\\
Figure~\ref{kepvort_noenergy} shows the pressure in the simulation with HLLC fluxes without the modified gravitational energy transport scheme. A strong numerical instability emerges already after two dynamical time scales, shortly after the inner spiral arm hits the inner boundary, The instability grows quickly and results in an abort of the simulation, since the internal energy becomes negative. We observe this instability arising in simulations with angular momentum conservation independently of any other parameters or a central point mass potential, once a non-axisymmetric perturbation hits the inner boundary. Additionally its occurrence is independent from the chosen boundary condition.
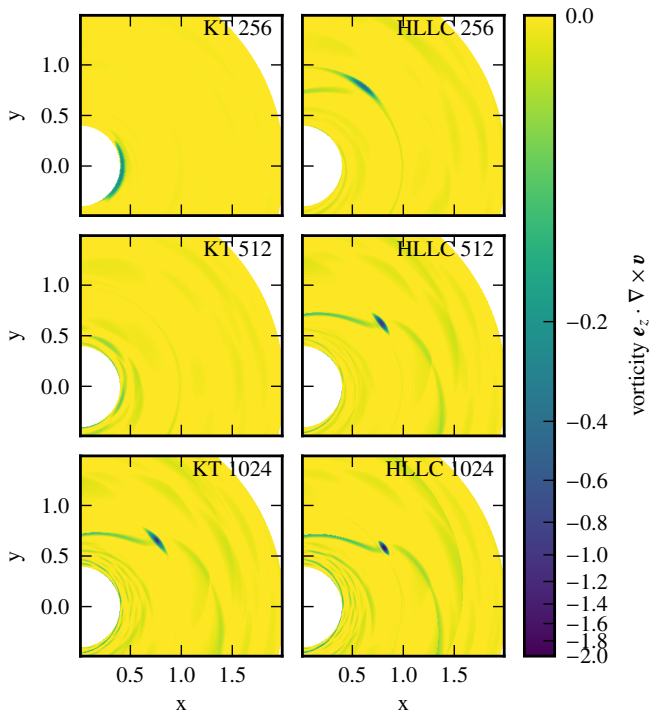
\figpgf{sims/figure/keplerianvortex/}{vorticitygrid_0030}{kepvort}{Vorticity of the vortex transport in a disk with keplerian rotation after $30$ orbits for KT and HLLC fluxes. The colormap is scaled with the $\arcsinh$ function, thus linear near zero and otherwise logarithmically. This enables us to observe the small details in the vorticity. The coarse resolutions using the KT fluxes cannot preserve the vortex for $30$ orbits. The simulation with the finest resolution HLLC $1024$ shows a clear improvement over the already usable KT $1024$ and HLLC $512$ simulations.}
Figure~\ref{kepvort} shows the vorticity of the vortex after $30$ orbits at different resolutions. If the KT fluxes are used, the vortex is already completely destroyed at this time for the radial resolutions of $256$ and $512$. Only a radial resolution of $1024$, which is nearly inaccessible (by cpu time) in realistic, often restarted simulations, the vortex can be preserved for longer time. The simulations with HLLC fluxes show a concise anticyclonic vortex, which is to some extent already at the lowest resolution preserved. Since the start of the simulation the vortex is sheared out by the differential rotation. A inner density spiral wave can still be observed. Residuals of the outer density spiral wave, as well as its numerous reflections at the outer boundary can be adumbrated. The simulations with higher resolutions show that this one is not converged yet. A resolution of $512$ shows a similar result as the KT fluxes at $1024$. The HLLC $1024$ simulation raises the overall definition again. The vortex size diminishes and in particular the outer spiral arm is more pronounced.\\
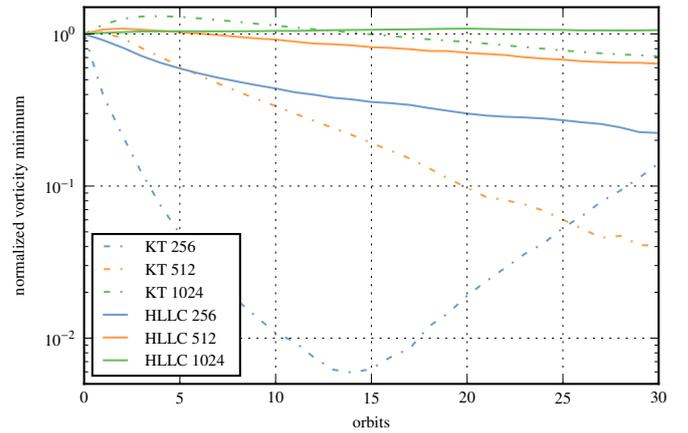
\figpgf{sims/figure/keplerianvortex/}{minvort}{kepvort_minvort}{Progress of the normalized vorticity minimum for the different simulations of the vortex transport in a disk with keplerian rotation. At all resolutions the HLLC fluxes show much smaller decay rates then the KT fluxes. Also their decay rate declines, while it stays nearly constant for the KT fluxes over the total simulation time.}
Figure~\ref{kepvort_minvort} shows the progress of the vorticity minimum for simulations of different resolutions using KT and HLLC fluxes. Figure~\ref{kepvort} shows that after $30$ orbits the vortex is completely dissolved using KT fluxes and resolutions of $256$ and $512$. We can therefore assert that vorticies with vorticities of less than $10\%$ of their initial value, are completely dissolved. The rise in vorticity in the simulation with the coarsest resolution can be explained by accumulation of vorticity at the inner boundary, but is not related to the conservation of the vortex. The simulation with HLLC fluxes at the smallest resolution has still about $20\%$ of its initial vorticity left in the vortex center. Although the decrease in vorticity is less pronounced compared to the simulations with KT fluxes, low spatial resolution leads to
a decay of the vortex likewise. This is confirmed by its change of position to smaller radii. The KT fluxes at $1024$ and the HLLC fluxes at $512$ indeed show a similar vorticity at the end of the simulation. Clearly this is the result of a switch on effect for the KT fluxes, which raises the initial vorticity at first. Afterwards the decay rate is greater in the KT fluxes simulation. The simulation with HLLC fluxes and $1024$ radial cells preserves the vortex pretty well. The vorticity raises a little bit, which can be explained by a change in the vorticity profile inside of the vortex. Of course the initial conditions do not exactly represent the by the flow adjusted solution in this setup. Overall we can note that the KT fluxes show a exponential decay at a constant rate in each simulation. In contrast the HLLC fluxes show for small resolutions a slow fading of the much weaker decay rate.\\
The simulations in this paper using HLLC fluxes can be compared to the FARGO simulations in \cite{2012A&A...545A.152M}. The flow in simulations with a resolution of $1024\times 4096$ cells show similar vortex shape and density spiral waves in both cases. The progress of the minimal vorticity is for all resolutions in very good agreement.\\
We can show that the vortex is dissolved within a short time, if one does not use HLLC fluxes \citep{Toro1994} (in contrast to the HLL \citep{Harten1983}, which are very similar to the KT fluxes) and the modified formulation of the energy equation. This shows the superior definition of a HLLC solver especially for vortex flows in combination with exact angular momentum conservation. 

\section{Conlusions}
In this paper we have introduced a new numerical algorithm to execute self-gravitating multi-scale simulations of barred galaxies with a particular interest in the black-hole accretion flow. Since angular momentum transport plays a crucial role for accretion disk dynamics, it is of great importance to reflect the angular momentum conservation in the numerical scheme. This is challenging in particular in a rotating frame of reference, since fictitious forces require careful handling. Therefore, we have introduced an implementation of a conservative numerical scheme for the solution of the Euler equations in a rotating frame of reference with angular momentum transport. This was combined with a Lagrangian advection scheme for the azimuthal transport and implemented in the astrophysical hydrodynamics software package \textsf{FOSITE}.\\
Furthermore we have implemented an efficient pseudo-spectral self-gravity solver, which is usable on polar grids with arbitrary radial scaling. The selection of a time-independent vertical disk structure allows by precalculation of a Fourier transformation to safe a considerable amount of computation time at each timestep.\\
At last we have introduced a modified implementation of gravitational source terms in the energy equation to avoid small but accumulating errors for the internal energy. In this method gravitational energy is advected with the flow and the gradient of the gravitational potential as source term can be avoided. The common implementation leads to non-axisymmetric instabilities arising at the inner boundary of the computational domain, especially if exact angular momentum conservation is enabled as well. We show that only the combination of angular momentum conservation, gravitational energy advection and contact resolving Riemann solvers (e.g. HLLC) is able to accurately resolve the important transport of an isentropic vortex in a Keplerian accretion disk.\\
In the second paper \citep{Jung2017b} we will use this new numerical algorithm to examine the self-gravitating flow of gas from galaxy regions at large radii to the smallest scales, driven by a stationary central stellar bar in a rotating frame of reference, with precise measurements of the black-hole accretion rates. This analysis will include both, isothermal simulations and simulations including the energy equation and a simple cooling mechanism to allow for marginal stable accretion disk flows.

\bibliography{paper}

\begin{appendix}
\section{Euler equations in polar coordinates}\label{sec:polarcoo}
The equations \eqref{eq:pluto1} - \eqref{eq:pluto3} in polar coordinates with arbitrary radial scaling are:
\begin{alignat}{3}
\frac{\partial \rho}{\partial t} &+ \nabla \cdot \left(\rho \bm{u}\right) &&+ \bm{w}\cdot\nabla\rho &&= 0\label{eq:plutopolar1}\\
\pdiff{\left(\rho u_\xi\right)}{t} &+ \nabla\cdot\left(\rho u_\xi \bm{u} + p \bm{e}_\xi\right) &&+ \bm{w}\cdot\left(\rho u_\xi\right) &&= S_{m_\xi}\\
\pdiff{\left(\rho u_\phi\right)}{t} &+ \nabla_R \cdot  \left(\rho u_\phi \bm{u} + p \bm{e}_\phi \right) &&+ \bm{w} \cdot \nabla \left(\rho \bm{u}_\phi\right) &&= S_{m_\phi'}\\
\frac{\partial E'}{\partial t} &+ \nabla \cdot \left(\left(E'+p\right)\bm{u}\right) &&+ \bm{w}\cdot\nabla E' &&= S_{E'}.\label{eq:plutopolar4}
\end{alignat}
The sources are expanded so that they can be easily implemented in dependence of the physical fluxes:
\begin{align}
S_{m_\xi} &= \rho v_\phi^2 c_{\phi\xi\phi}\\
S_{m_\phi'} &= -\nabla_R\cdot\left(\rho w \bm{u}\right) + w \nabla\cdot\left(\rho\bm{u}\right)\\
S_{E'} &= \nabla\cdot\left(\frac{w^2}{2}\rho\bm{u}+\bm{w}\cdot\left(\rho\bm{u}\otimes\bm{u}+\mathbb{I}p\right)\right) -\frac{w^2}{2}\nabla\cdot\left(\rho \bm{u}\right)\nonumber\\
&\phantom{{}=} -w\nabla_R\cdot\left(\rho u_\phi \bm{u}\right) - w\nabla_R\left(\rho w \bm{u}\right).
\end{align}
The metric coefficients for polar coordinates using a monotonic differentiable radial scaling function $r\left(\xi\right)$ are:
\begin{align}
h_\xi &= \left| r'\left(\xi\right)\right|\\
h_\phi &= \left|r\left(\xi\right)\right|
\end{align}

\section{Modified Bessel functions}\label{sec:besselk0}
The modified Bessel function of the first and second kind are \citep{1965hmfw.book.....A}:
\begin{align}
I_\alpha\left(x\right) &= \sum\nolimits_{m=0}^\infty \frac{1}{m!\, \Gamma\left(m+\alpha+1\right)}\left(\frac{x}{2}\right)^{2m+\alpha}\\
K_\alpha\left(x\right) &=\frac{\pi}{2}\frac{I_{-\alpha}\left(x\right)-I_\alpha\left(x\right)}{\sin\left(\alpha\pi\right)}.
\end{align}

\section{Fourier transformation}\label{sec:fouriertrafo}
Since a number of different scalings for the Fourier transformation and convolutions are common, we note the in this work used notation and abbreviations: .
\begin{align}
\hat{f} &:= \mathcal{F}\left(f\right)\\
\hat{f}_m &= \frac{1}{2\pi}\int_0^{2\pi}f\left(\phi\right)e^{-im\phi}\dd\phi\\
\mathcal{F}\left(f\star g\right) &= 2\pi \mathcal{F}\left(f\right)\cdot \mathcal{F}\left(g\right)
\end{align}

\end{appendix}
\end{document}